%% file: IEEE-conference-template-062824.tex
\documentclass[conference]{IEEEtran}
\IEEEoverridecommandlockouts

\usepackage{cite}
\usepackage{amsmath,amssymb,amsfonts}
\usepackage{algorithmic}
\usepackage{graphicx}
\usepackage{textcomp}
\usepackage{booktabs}
\usepackage{tabularx}
\usepackage{multirow}
\usepackage{xcolor}
\usepackage{csquotes}
\usepackage[normalem]{ulem}
    
\usepackage{glossaries}
\usepackage{url}
\usepackage{hyperref}
\input{glossary}

\def\BibTeX{{\rm B\kern-.05em{\sc i\kern-.025em b}\kern-.08em
    T\kern-.1667em\lower.7ex\hbox{E}\kern-.125emX}}

\usepackage{eso-pic}
\begin{document}
\AddToShipoutPictureFG*{%
  \AtPageUpperLeft{%
    \raisebox{-1.0cm}{%
      \makebox[\paperwidth][c]{%
        \parbox{0.9\paperwidth}{%
          \centering
          \small\itshape
          Preprint version of a manuscript accepted for publication in the
          \textit{37th IEEE International Symposium on Software Reliability
          Engineering (ISSRE) 2026}, Limassol, Cyprus.
        }%
      }%
    }%
  }%
}

\title{Understanding Online Failure Prediction in Linux Through Complementary Multi-View Explainability}

\author{\IEEEauthorblockN{Diogo Dória}
\IEEEauthorblockA{\textit{University of Coimbra, CISUC/LASI} \\
\textit{Department of Informatics Engineering}\\
Coimbra, Portugal \\
dbdorio@student.dei.uc.pt}
\and
\IEEEauthorblockN{João R. Campos}
\IEEEauthorblockA{\textit{University of Coimbra, CISUC/LASI} \\
\textit{Department of Informatics Engineering}\\
Coimbra, Portugal \\
jrcampos@dei.uc.pt}
}

\maketitle

\begin{abstract}
Accurate \gls{OFP} has been shown to be feasible in \glspl{OS} settings, but prediction alone is not sufficient for practical adoption. Without diagnostic insight, operators have limited basis to trust alerts or decide how to respond. Moreover, even when predictive accuracy is high, it is often unclear whether models are capturing meaningful failure processes or merely exploiting workload-specific noise and incidental correlations in telemetry. This paper reports a practical experience building and evaluating an explainable \gls{OFP} pipeline for Linux \glspl{OS}. We combine consensus-based feature selection for detection with temporal onset analysis, subsystem-level causal analysis, and complementary diagnostic mechanisms to support failure interpretation. Evaluated under strict cross-workload conditions with frozen training artifacts, it achieved 91-94\% detection on unseen workloads without retraining, while maintaining false alarm rates below 1\%. However, failure mode diagnosis proved substantially more sensitive to workload shift, and several diagnostics mechanisms showed limited effectiveness for specific failure types. Our experience highlights three main lessons: i) detection generalizes more robustly than diagnosis across workload changes; ii) early-warning capability depends strongly on the failure mode, ranging from 38 to 215 seconds in our study; and iii) unseen failure modes are not reliably diagnosable from related training modes alone, providing 0\% accuracy under Leave-One-Mode-Out (LOMO) evaluation. Taken together, these results show the value of complementary explainability mechanisms for interpreting accurate failure predictions, revealing when predictive signals reflect transferable failure structure and when diagnostic generalization breaks down under workload variation. 
\end{abstract}
\begin{IEEEkeywords}
Online Failure Prediction, Explainable AI, Linux, Root Cause Analysis, Failure Diagnosis.
\end{IEEEkeywords}

\glsresetall

\section{Introduction}

Unexpected system failures remain persistent threats to operational reliability in mission-critical domains. In healthcare infrastructure, financial services, and cloud platforms, failures translate directly into service outages, data loss, and economic damage \cite{b7}. Traditional monitoring approaches are fundamentally reactive, triggering alerts only after anomalies cross predefined thresholds or services have visibly degraded \cite{b7,b8}. By the time alerts fire, underlying failure processes may have progressed beyond recovery. 

\gls{OFP} represents a paradigm shift toward anticipatory system management \cite{b7,b8}. Rather than waiting for severe symptoms, OFP systems continuously analyze high-frequency telemetry to detect failure precursors while intervention remains possible, providing early-warning windows that enable automated mitigation before catastrophic collapse.

Despite significant research progress, practical \gls{OFP} deployment remains constrained by three interrelated challenges. First, interpretability remains limited. Modern \gls{OFP} systems can achieve strong predictive accuracy \cite{b1,b3}, but often without exposing why failures are predicted, which subsystems are involved, or how degradation propagates. As a result, operators have limited basis to trust alerts or decide how to respond. More fundamentally, high predictive performance alone does not reveal whether models are capturing meaningful failure processes or merely exploiting workload-specific noise and incidental correlations in telemetry. Second, failures unfold through multi-stage temporal interactions involving memory, disk I/O, CPU, network, and kernel subsystems. Understanding when reliable warning signals emerge, and which subsystems act as drivers versus downstream recipients, requires explicit temporal and causal analysis beyond point-wise anomaly scoring. Third, generalization across operational contexts remains uncertain: telemetry patterns under CPU-intensive execution differ fundamentally from I/O-heavy or memory-bound workloads, and most approaches are evaluated under training-like conditions, leaving open whether learned signatures remain valid when deployment contexts diverge \cite{b8}. 

These challenges are interdependent. Interpretability requires linking predictions to subsystem behavior and plausible failure dynamics; temporal characterization requires distinguishing persistent precursors from transient fluctuations; and generalization requires identifying signals that reflect transferable degradation structure rather than workload-specific artifacts. Addressing these questions therefore benefits from combining accurate prediction with complementary analyses for feature selection, temporal characterization, causal interpretation, and diagnosis.

This paper reports practical experience building and evaluating an explainable \gls{OFP} pipeline for Linux systems. The goal is to complement failure prediction with additional analyses that help interpret alerts, characterize failure evolution, and assess whether the learned signals reflect meaningful failure structure. In practical terms, these analyses are intended to support operators and reliability engineers after an alarm by helping them assess its credibility, understand likely subsystem involvement, distinguish among known failure modes, and judge whether any plausible intervention is suggested. To this end, we combine consensus-based feature selection for detection with temporal progression analysis, subsystem-level causal inference, and layered diagnostic mechanisms. The approach is evaluated under strict cross-workload conditions using frozen artifacts, so that observed generalization reflects robustness of the learned signals rather than adaptation to the validation workloads.

The main contributions of this paper are:
\begin{itemize}
    \item A \textbf{multi-view consensus methodology for cross-workload robust feature selection}, synthesizing statistical deviation, supervised importance, and clustering-based discrimination into compact, mode-specific feature sets for \gls{OFP}.
    \item \textbf{Temporal and causal characterization of failure evolution} through \textbf{\gls{ERS}}, which quantifies population-level early-warning potential per failure mode, and \textbf{subsystem-level Granger causality} analysis, which exposes directed propagation pathways distinguishing upstream drivers from downstream symptoms.
    \item \textbf{Layered diagnostic explainability through contrastive fingerprints} for pairwise mode discrimination, \textbf{symbolic rule-based diagnosis} for human-readable patterns, and \textbf{counterfactual interventions for prescriptive guidance}, providing complementary views for failure interpretation and diagnosis.
    \item A practical \textbf{evaluation of the proposed methods in a Linux-based setting} under strict cross-workload validation with frozen artifacts, showing that detection generalizes more robustly than diagnosis, that early-warning capability is strongly failure-mode dependent, and that previously unseen failure modes are not reliably diagnosable from related training modes alone.
\end{itemize}

The remainder of this paper is organized as follows. Section II reviews background and related work. Section III presents the pipeline design and complementary analyses. Section IV describes the experimental setup. Section V reports empirical results. Section VI discusses findings, representativeness, and limitations. Section VII concludes.

\section{Background and Related Work}

This section introduces relevant background concepts and related work. 

\subsection{\acrfull{OFP}}
The fault-error-failure chain describes how undesired events propagate in computing systems: a fault activates under specific triggers, producing an internal error that may propagate to external failure \cite{b7}. \gls{OFP} operationalizes failure prediction by continuously analyzing runtime telemetry to anticipate failures before service disruption \cite{b7,b8}. As a surveyed comprehensively by Salfner et al. \cite{salfner2010}, OFP methods rely on both historical data, used to train predictive models, and the current state of the system, captured through online monitoring. A prediction made at time $t$ targets a future interval beginning at $t+\Delta t_l$ and lasting for $\Delta t_p$, where $\Delta t_l$ and $\Delta t_p$ denote the \textit{lead time} and \textit{prediction window}, respectively. Thus, the model estimates whether a failure will occur in the interval $[t+\Delta t_l,\; t+\Delta t_l+\Delta t_p]$. Unlike traditional reliability models based mainly on historical statistics, \gls{OFP} incorporates current observations to estimate failure likelihood within that prediction window, enabling corrective action before collapse.

Machine learning has become the dominant paradigm for OFP. Ensemble methods combining Random Forest or gradient-boosted trees report detection rates of 85-90\% on fault-injection datasets \cite{b1,b3}, while deep learning approaches operating on logs or multivariate telemetry achieve comparable or slightly higher accuracy (up to 92\%) under controlled conditions \cite{b2,b4}. However, most systems focus on binary detection (failure versus healthy) rather than diagnostic identification of failure modes. Recent evaluations reveal that model performance degrades significantly when failure rates drop below 5\% \cite{b2}, highlighting the difficulty of learning reliable predictive signals under class imbalance.

A persistent challenge in \gls{OFP} research is the availability of suitable failure data. OFP has been studied across diverse domains, including hard drive failure prediction using SMART attributes \cite{zhang2020minority}, optical network degradation \cite{wang2017failure}, and job failure prediction in cloud computing environments using publicly available traces such as Google cluster data \cite{chen2014failure,jassas2018}. However, the most detailed measurements in such datasets are typically available at the level of jobs, tasks, or containers rather than the OS as a whole, and often at coarser temporal granularity (e.g., five-minute intervals). As a result, they are less suitable for studying short-horizon OFP with activation-aligned degradation analysis and subsystem-level failure interpretation in a full-fledged OS. Because real OS-level failure data are scarce, inaccessible, or unavailable for new systems, controlled fault injection, particularly Software-Implemented Fault Injection (SWIFI), has become an accepted practical alternative for generating realistic labeled failure data under reproducible conditions \cite{b7,b8,b9,campos2020}.

Another persistent limitation in the OFP literature is evaluation methodology. Most studies partition data randomly within a single workload, assuming distributional continuity between training and test sets. This assumption is systematically violated when workloads change, since transitions from CPU-bound to I/O-bound execution can fundamentally shift telemetry distributions. Cross-workload evaluation, where models are trained on one workload and tested on operationally distinct workloads without retraining, provides a substantially more demanding and realistic assessment of generalization. Yet such evaluation remains rare in OFP research.

Furthermore, few studies systematically evaluate cross-mode transferability, that is, whether failure signatures learned for one set of failure modes generalize to unseen failure mechanisms. Understanding where generalization holds and where it breaks is essential for principled deployment but is rarely addressed empirically.


Despite advances in predictive accuracy, a critical gap remains between detection capability and diagnostic utility. Prediction accuracy has advanced faster than diagnostic transparency, limiting trust and actionable use in high-stakes environments. More importantly, high predictive performance alone does not establish that a model has learned meaningful failure dynamics. In the absence of interpretive analysis, a predictor may achieve strong results by exploiting workload-specific noise, incidental correlations, or other spurious telemetry patterns rather than signals that genuinely reflect failure processes. As a result, understanding and interpreting what \gls{OFP} models are actually learning remains comparatively underexplored.

\subsection{Explainability in System Reliability}

Feature attribution methods, particularly SHAP and LIME, dominate current \gls{XAI} research. Comprehensive reviews \cite{b16,b17} document widespread adoption across domains, but also identify temporal stability as a significant challenge: explanations that fluctuate rapidly over successive observations can undermine operator trust. Standard \gls{XAI} methods assume static prediction scenarios and rarely address temporal continuity, limiting applicability to online monitoring. 

Causal analysis offers a complementary perspective. Granger causality provides a framework for inferring directed relationships based on temporal precedence without requiring prior structural knowledge. Recent applications in industrial fault diagnosis demonstrate utility for root cause analysis and propagation path identification \cite{b18,b19}. However, integration of causal inference into online prediction pipelines remains limited, most applications are retrospective rather than embedded in streaming detection workflows.

Counterfactual reasoning \cite{b20} identifies minimal input changes that would alter predictions, offering actionable diagnostic guidance. However, applying counterfactual methods to temporal monitoring contexts, where plausibility constraints must respect system dynamics, remains largely unexplored.

Prior work has explored these elements individually, but their combined use as complementary analyses for interpreting \gls{OFP} remains limited.

\section{Experimental Methodology}

The proposed framework transforms high-frequency \gls{OS} telemetry into actionable failure insight through an integrated pipeline combining multi-view feature analysis, temporal-causal modeling, and layered diagnostic explainability. Rather than treating failure prediction as an isolated binary classification task, the design explicitly models symptom persistence, feature robustness across analytical perspectives, and directed temporal dependencies among subsystem-level dynamics. 

\begin{figure}[htbp]
    \centering
    \includegraphics[width=\columnwidth]{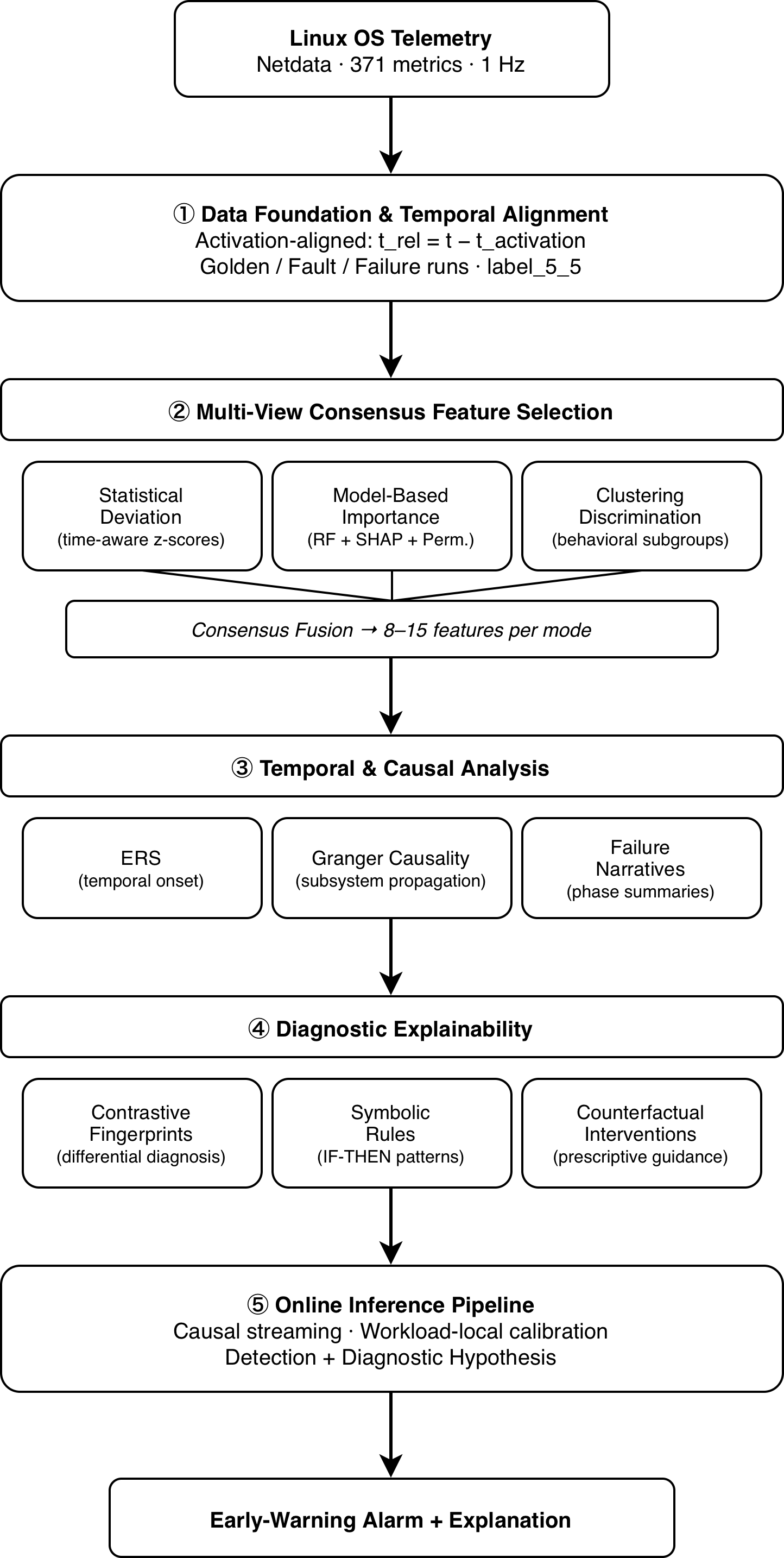}
    \caption{Framework overview diagram}
    \label{fig:diagram}
\end{figure}

The framework operates in five sequential stages: (1) Data foundation and temporal alignment, (2) multi-view consensus feature selection, (3) temporal and causal analysis, (4) diagnostic explainability mechanisms, and (5) online inference. Each stage addresses a specific analytical objective while contributing to a coherent end-to-end prediction and diagnosis process. A key design principle is that no single analytical perspective, statistical deviation, model importance, or structural clustering captures failure complexity alone; consensus across views provides robustness against method-specific biases. Together, these layers are intended not as competing alternatives, but as complementary tools for detection, interpretation, and diagnosis.

\subsection{Study Context and Data Foundation}


Much of the existing \gls{OFP} literature targets narrower settings, individual components such as hard drives \cite{zhang2020minority}, systems with well-defined quality indicators \cite{wang2017failure}, or large-scale traces such as Google cluster data for job-failure prediction \cite{chen2014failure}. These sources measure at the level of jobs, tasks, or containers rather than the \gls{OS} as a whole, and often at coarser granularity (e.g., five-minute intervals), making them less suitable for short-horizon, activation-aligned, subsystem-level \gls{OFP}. Controlled fault injection is therefore our practical means of obtaining realistic, labeled, and reproducible \gls{OS}-level failure data.

In this work we use a Linux \gls{OS} failure dataset publicly available, generated through fault injection \cite{campos2020}. Linux provides a practically relevant setting for such analysis given its widespread use across diverse application domains. Faults were injected at the kernel binary level during controlled workload execution, emulating realistic low-level programming errors including control-flow corruption, pointer faults, memory allocation errors, and omission faults. System behavior was monitored using Netdata at one-second resolution, producing 371 synchronized telemetry metrics per timestamp. These metrics are organized into five subsystem categories: CPU and process scheduling, memory management, disk and I/O, network, and kernel internals. Each experiment run is categorized as a golden run (no fault injection, healthy baseline), a fault run (fault injected, no failure observed), or a failure run (fault propagates to an observable system failure). The dataset comprises 43 golden runs, 146 fault runs, and 42 failure runs. Failure runs are distributed across five failure modes derived from the outcome and log-based failure categories reported in \cite{campos2020}: \textit{Crash} ($n=6$), for system crashes; \textit{Hang} ($n=3$), for system hangs; and three non-fail-stop categories inferred from system logs according to their likely root cause: \textit{CPU} ($n=11$), \textit{Memory} ($n=16$), and \textit{Kernel} ($n=6$).

Three operationally distinct workloads are used to study cross-workload generalization. A CPU-intensive workload is used for training and for deriving all learned artifacts, including consensus features, thresholds, narratives, and diagnostic mechanisms. Two unseen workloads are then used for validation: NGio, which stresses disk and file-system subsystems through sustained buffer cache synchronization and writeback activity, and NGmtr, which exercises memory bandwidth and cache hierarchy through multi-threaded matrix computations. These two validation workloads contribute 45 (NGio) and 50 (NGmtr) failure runs, respectively. Together, these workloads induce two forms of distribution shift: a change in dominant resource bottleneck (CPU$\rightarrow$I/O) and a change in execution characteristics within computation-heavy workloads. All workloads share the same fault injection mechanism, fault taxonomy, monitoring system, and failure labeling criteria; the workload profile is the main source of operational variation. The workload executed for 600s (unless a failure abruptly terminated the execution). 

A key methodological choice is to align all analyses to the moment of fault activation rather than to failure occurrence. For each run, a relative time index $t_{rel}=t-t_{activation}$ is defined, where $t_{rel}=0$ corresponds to the onset of potential degradation. This activation-aligned coordinate system enables consistent comparison of early, intermediate, and late failure dynamics across runs with heterogeneous durations and failure latencies. Two temporal windows are considered: the full degradation window from activation to failure, and a fixed 120-second horizon for prospective early-warning analysis.

Telemetry samples are annotated according to whether failure is expected within a short prediction horizon. Throughout the study, we use a five-second lead time and a five-second prediction window, a configuration chosen to retain actionable anticipation while avoiding labels that are either too close to failure to be useful or too far in advance to remain reliable.

\subsection{Multi-View Consensus Feature Selection}

Reliable failure indicators should remain informative across different analytical viewpoints. A feature that appears important under only one method may reflect artifacts of that method's assumptions. The consensus layer integrates three complementary views to identify features with multi-perspective support. 

The first view, statistical deviation (time-aware ranking), evaluates each telemetry metric by how strongly and consistently it diverges from a healthy baseline after fault activation. A golden run baseline provides per-second reference distributions (mean and standard deviation) for each feature. Deviation is quantified using absolute z-scores: 

\begin{equation}
\label{eq:zscore}
z_{t,f} = \frac{\left| x_{t,f} - \mu^{(G)}_{t,f} \right|}{\sigma^{(G)}_{t,f} + \epsilon}
\end{equation}

Scores are aggregated per feature using the median (typical deviation) and 95th percentile (transient spikes). To favor failure mode-specific symptoms over generic stress indicators, each feature's deviation under each failure mode is contrasted against its deviation under all other failure modes through a specificity adjusted scoring function that rewards features whose deviation is both strong and more pronounced in the target mode. 

The second view, model-based importance, derives features importances from supervised models trained to discriminate failure progression from healthy behavior on activation-aligned windows. Random Forest (RF), which was shown to achieve the good predictive performance in previous works, serves as the primary model with importance extracted through three complementary mechanisms: SHAP values (local contribution aggregated to global relevance), permutation importance (performance impact under feature shuffling), and native impurity-based importance. A blended importance score combines these signals with stability-weighted aggregation across Group-KFold cross-validation folds, where all samples from a run occupy the same fold to prevent temporal leakage. Elastic-Net Logistic Regression and XGBoost serve as secondary benchmarks, chosen because they differ structurally from RF in how they assign feature importance, Elastic-Net uses coefficient magnitude under regularization, while XGBoost uses gradient-based splits. This diversity ensures that retained features are not artifacts of a single model family's inductive bias.

The third view, clustering-based discrimination, captures behavioral heterogeneity within the same failure mode by identifying recurring degradation subgroups across runs. Each run is represented by a fixed-length descriptor constructed by dividing the post-activation window into five equal temporal segments, a partition chosen to balance temporal resolution against descriptor dimensionality, ensuring enough segments to capture early, intermediate, and late degradation phases while keeping the run-level representation compact enough for stable clustering with limited sample sizes, and computing the mean, standard deviation, and slope of each consensus feature within each segment. This produces a 15-dimensional vector per feature per run, capturing both the level and temporal trend of degradation. For modes where consensus clustering identifies stable multi-cluster structure, discriminative features are extracted using the Kruskal-Wallis test with Benjamini-Hochberg False Discovery Rate (FDR) correction \cite{benjamini1995}. For modes where clustering does not reveal meaningful subgroups (i.e., a single-cluster solution is preferred), this view contributes no additional discriminative features, and the consensus fusion proceeds using the remaining two views.

Finally, consensus fusion aggregates the three view-level rankings for each failure mode. Rankings are converted to normalized scores via inverse-rank transformation, and a consensus score is computed using equal weighting across views. Correlation pruning ($|r|>0.90$ threshold) removes redundant features. This threshold retains features that share up to 80\% of variance while removing near-duplicates that would inflate the consensus ranking without adding independent information. This provides compact features sets (8-15 features per mode from 371 candidates) that form the basis for all downstream analysis.

\subsection{Temporal and Causal Analysis}

The consensus stage identifies the features that are most relevant for distinguishing failure-related behavior from healthy behavior. Temporal and causal analysis then addresses two complementary questions: when do reliable warning signals emerge, and how does degradation propagate across subsystems?

\gls{ERS} quantifies the temporal onset of detectable instability across the set of available failure runs for each mode, rather than for individual runs in isolation. Operating in a failure-aligned coordinate system ($t_{rel} = t -t_{fail}$, where $t_{rel}=0$ marks entry into the prediction window), each consensus feature is converted to standardized z-scores against a golden-run baseline. A feature is marked unstable at time t if $|z_{t,f}|\geq 2.0$ conventional two-sigma abnormality criterion that separates baseline variation from sustained deviation at the 1Hz telemetry resolution used in this study. Three run-level indicators are computed per timestep: unstable feature count (scale), mean $|z|$ over unstable features (intensity), and rapid-change count over a sliding window (volatility). 

The run-level ERS onset time is defined as the earliest $t_{rel}$ where the unstable feature count exceeds an adaptive threshold $\theta=max(\alpha \times max_t(unstable\_count(t)), C_{min})$, with $\alpha=0.35$ and $C_{min}=5$. The scaling factor of $\alpha=0.35$ was chosen empirically to require that at least $35\%$ of the peak instability observed in a run must be reached before onset is declared, avoiding premature triggering from transient fluctuations. The floor $C_{min} = 5$ ensures that onset requires a minimum of five simultaneously unstable features, preventing false onset declarations in runs with low overall instability. Mode-level summaries report median and inter-quartile range of onset times, enabling rigorous cross-mode comparison of early-warning potential. Each feature is also mapped to a subsystem category (CPU, memory, disk, network, kernel), providing subsystem-level onset profiles. 

Granger causality \cite{granger} tests whether past values of one time series improve prediction of another beyond the target's own history. Here, it is applied at the subsystem level to infer directed progression graphs during post-activation evolution. For each failure mode and ordered subsystem pair $X\xrightarrow{}Y$, Granger tests are performed up to a maximum lag of 8 second. This cap reflects 1Hz sampling rate and focuses the analysis on short-range causal propagation, while avoiding unstable parameter estimation under longer lags given the limited post-activation observation windows. An edge $X\xrightarrow{}Y$ is considered significant in a given run if the F-test rejects the null hypothesis that past values of $X$ do not improve prediction of $Y$ ($p<0.05$). Evidence is aggregated across runs, only edges that are significant in at least $50\%$ of runs for a given failure mode are retained, ensuring that reported causal relationships are reproducible across the majority of observed failure instances rather than driven by isolated runs. For robustness analysis, we additionally label an edge stable if it is significant in more than $80\%$ of a mode's runs, this stricter criterion is used only to assess reproducibility and is not the edge-retention threshold.

\begin{equation}
\label{eq:granger}
W_{X \rightarrow Y} = \mathrm{FracSig}_{X \rightarrow Y} \cdot \left( -\log \tilde{p}_{X \rightarrow Y} \right)
\end{equation}

where $FracSig_{X\rightarrow Y}$ is the fraction of runs in which the Granger F-test for the edge $X\rightarrow Y$ is significant ($p<0.05$), and  $\tilde{p}_{X\rightarrow Y} $ is the median p-value of the F-test across runs. Subsystems with high outgoing strength act as upstream drivers, those with high incoming strength act as downstream recipients. The resulting mode-specific directed graphs provide compact propagation signatures that complement the temporal onset profiles from ERS. 

Failure Narratives are built using consensus features with reliable ERS latencies that are grouped into three progression phases (early, mid, late) based on latency quantiles. Within each phase, representative features are selected using consensus score as the primary criterion and tagged by subsystem, producing phase-structured summaries of how each failure mode unfolds temporally. 

\subsection{Diagnostic Explainability Mechanisms}

Beyond detection and temporal characterization, operational deployment requires differential diagnosis (which failure mode?) and intervention guidance (what should change?). Three complementary mechanisms address distinct diagnostic questions.

Contrastive fingerprints were designed using each unordered failure mode pair ($m_A, m_B$), a Logistic Regression model with Elastic-Net regularization is fit on activation-aligned consensus features. Signed coefficients identify which features most strongly separate the two modes: positive weights favor one mode, negative weights the other. Group-level permutation testing ($B=500$ permutations with run-level label shuffling) assesses statistical significance, and group-aware cross-validation (GroupKFold by run) prevents temporal leakage. Answering the question: What distinguishes mode A from mode B?

Hybrid Rule-based diagnosis was implemented for each mode, a one-vs-rest rule extraction pipeline produces compact IF-THEN rules from activation-aligned telemetry. Run-level descriptors encode pre/post-activation mean, change magnitude, ratio and constancy indicators (stability-to-instability transitions). Shallow decision trees under bagging generate candidate conjunctive rules, which are filtered by support and purity and audited on the full run set. Rules can be evaluated with the earliest firing time providing an interpretable early/late indicator. This mechanism answers to what human-readable conditions characterize this failure mode?

Counterfactual analysis \cite{b20} seeks to identify minimal, bounded feature changes that would move a failure indicative state toward healthy classification. Plausibility constraints derived from golden run bounds (5th-95th percentiles) prevent unrealistic recommendations. A lightweight Logistic Regression surrogate provides a tractable decision surface and $l_1$-regularized optimization promotes sparse edits: 

\begin{equation}
\label{eq:counterfactual}
\min_{\Delta x} \; \lVert \Delta x \rVert_1
\quad \text{s.t.} \quad
P(\text{Failure} \mid x_0 + \Delta x) < \tau
\end{equation}

where $\Delta x =$ vector of feature edits to be optimized, $x_0 =$ high confidence failure-like starting point, $P(\text{Failure} \mid x_0 + \Delta x) = $ surrogate failure probability after the edit and $\tau =$ safety threshold ($0.3$) is a safety threshold chosen conservatively below default decision boundary of 0.5 to ensure that counterfactual states are moved well into the healthy region rather than merely crossing the boundary. Counterfactual answers what would need to change to prevent this failure classification?

These three mechanisms are designed to be complementary rather than competing. Contrastive fingerprints provide quantitative pairwise discrimination, symbolic rules provide auditable global patterns, and counterfactual edits provide instance-specific prescriptive guidance.

\subsection{Online Inference Pipeline}

The framework operates under strict causal streaming constraints: at each timestep $t$, only observations from times $\leq t$ are available. Raw feature values are standardized using workload-local golden-run statistics to account for distributional differences across operational regimes while preserving learned feature sets and weights. For each failure mode m, a weighted deviation score aggregates absolute standardized deviations across its consensus features: 

\begin{equation}
\label{eq:detection}
S(t) = \max_{m} \left( \frac{1}{W_m} \sum_{f \in F_m} \left| z_f(t) \right| \cdot w_{f,m} \right)
\end{equation}

The global detection score takes the maximum across all modeled failure modes, following the principle that failure is anticipated when any mode exhibits sufficiently strong deviation. Alarm thresholds are calibrated per workload using the 99.5th percentile of golden run score distributions, targeting a nominal 0.5\% false alarm rate. Persistence filtering (K=5 consecutive seconds above threshold) suppresses transient false alarms. The mode achieving the maximum score at alarm time provides a preliminary diagnostic hypothesis, which can be refined by the diagnostic mechanisms described above. 

\subsection{Evaluation Protocol}

The evaluation protocol is designed to reflect realistic deployment conditions. All learned artifacts from the training phase remain frozen during validation: no retraining, feature reselection, or model adaptation is performed. Only alarm thresholds are recalibrated per workload using golden run score distributions, isolating distributional scale differences while preserving all learned representations.

All detectors operate under causal constraints: at each timestep, only past observations are available. This prevents information leakage and ensures that measured lead times reflect genuine early-warning capability.

Performance is quantified using four metrics aligned with operational requirements: detection rate (proportion of failure runs with a pre-failure alarm), median lead time (alarm-to-failure interval), false alarm rate (proportion of golden runs with spurious alarms), and diagnostic accuracy (proportion of detected failures correctly classified).

Cross-mode transferability is assessed through Leave-One-Mode-Out (LOMO) analysis, which systematically withholds each failure mode from training and tests classification on the held-out mode, providing a direct assessment of how well learned signatures transfer to previously unseen failure mechanisms.

\section{Results}
This section presents the empirical evaluation of the proposed approach across three complementary dimensions: temporal-causal characterization of failure evolution, diagnostic explainability mechanisms, and cross-workload validation of generalization. The first two dimensions focus on the CPU-intensive training workload, which serves as the basis for developing and analyzing the learned artifacts in detail. The third then assesses how these artifacts transfer to the unseen NGio and NGmtr workloads under strict cross-workload conditions.

\subsection{Temporal-Causal Characterization}
On the CPU training workload, the consensus-based detector achieved 95.2\% detection rate with a median lead time of 82.5 seconds, while One-vs-Rest diagnosis reached 100\% mode classification accuracy among detected failure. Detection and diagnosis here are evaluated under Leave-One-Run-Out cross-validation (GroupKFold with 42 folds, each holding out a single run); all five failure modes are present in every training fold, so no mode is withheld at this stage; mode withholding occurs only in the separate LOMO analysis. These results confirm that the learned artifacts, consensus features, detection thresholds, and diagnostic models, are well-calibrated within training distribution.
The consensus layer produced compact, mode-specific feature sets from 371 candidate metrics. Across failure modes, $97$-$109$ features passed initial filtering, with the final consensus retaining approximately $26$-$29$\% of candidates per mode. Importantly, consensus integration shifted diagnostic signatures relative to single-view rankings. For example, while disk write activity was the primary statistical deviation for Crash failures, consensus prioritized system buffer metrics (system.ram.buffers) as the most robust cross-view signature, reflecting the system's buffering response rather than the triggering I/O stress. Similarly, kernel failures were uniquely characterized by inter-process communication (IPC) metrics (users.pipes), not network interrupt handlers as a single-view analysis would suggest.

Memory failures achieved the most compact consensus representation, converging strongly on kernel memory allocation metrics (mem.kernel.VmallocUsed). Hang failures, contrary to a scheduling latency hypothesis, were characterized by kernel stack and slab memory usage, indicating kernel-level resource exhaustion rather than CPU scheduling jitter. These results demonstrate that consensus integration mitigates single-view biases by filtering features with high deviation but low cross-perspective support. 

\begin{figure}[htbp]
    \centering
    \includegraphics[width=\columnwidth]{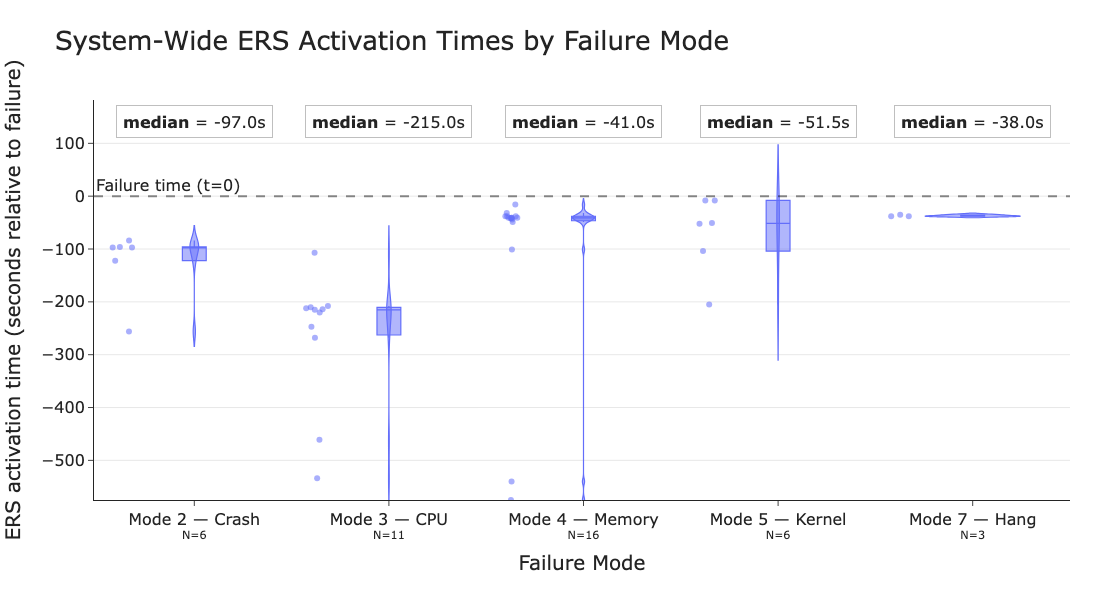}
    \caption{System-wide instability accumulation across failure modes.}
    \label{fig:ers_activation_fig}
\end{figure}

\gls{ERS} analysis, which quantifies when system-wide instability first becomes reliably detectable relative to failure, revealed a temporal hierarchy of early-warning capability that is fundamentally failure mode dependent. Table \ref{tab:ers_activation} reports median ERS onset times and lead time distributions for each failure mode. Figure \ref{fig:ers_activation_fig} visualizes the distribution of ERS activation times across failure modes, showing both the median onset and the inter-run variability for each mode.

\begin{table}[htbp]
\centering
\caption{ERS activation timing and early-warning potential}
\begin{tabular}{lcc}
\toprule
Failure Mode & Median Lead & IQR Range \\
             & Time (sec)  & (sec)     \\
\midrule
CPU          & 215         & 211--258  \\
Crash        & 97          & 96--116   \\
Kernel       & 51.5        & 19--91    \\
Memory       & 41          & 38--45    \\
Hang         & 38          & 36--38    \\
\bottomrule
\end{tabular}
\label{tab:ers_activation}
\end{table}

CPU failure exhibited the earliest and most robust detection (median $215$s, Inter-quantile Range (IQR): $211$-$258$s), consistently providing over three minutes of mitigation time through gradual resource exhaustion. Crash failures, contrary to the expectation of instantaneous faults, exhibited a robust signature with a compact distribution (median $97$s, IQR: $96$-$116$s). The narrow IQR suggests a deterministic precursor pattern offering a guaranteed warning window exceeding 90 seconds.

In contrast, Memory and Hang failures were characterized by late-onset, abrupt escalation. System-level statistical signatures remained indistinguishable from the baseline until a critical saturation tipping point, resulting in limited early-warning windows of 41 and 38 seconds, respectively. Kernel failures occupied an intermediate but highly variable position (median $51.5$s, IQR:$19$-$91$s), reflecting the heterogeneity of kernel-level fault-injection paths.

These results establish a key finding: early-warning capability is a property of the failure mechanism, not the detection algorithm. Failures with early-onset instability (CPU, Crash) afford substantial windows for automated remediation, whereas failures with late-onset signatures (Memory, Hang) necessitate rapid-response mechanisms.

Subsystem-level decomposition further clarified temporal onset patterns. Crash failures demonstrated a systemic shock pattern with instability detectable across CPU, Disk, and Memory subsystems as early as $-250$s before failure. Memory failures displayed synchronized saturation, with all subsystems remaining negligible until a simultaneous surge at approximately $-45$s. Hang failures were distinctively driven by the Disk/IO subsystem, validating an I/O-centric deadlock hypothesis.

\begin{table}[htbp]
\centering
\renewcommand{\arraystretch}{1.15}
\setlength{\tabcolsep}{3pt}
\footnotesize
\caption{Subsystem causal roles: out-strength (driver) vs in-strength (recipient). Values represent total aggregated causal weight from Granger Analysis.}
\begin{tabular}{l|cc|cc|cc|cc|cc}
\hline
\multirow{2}{*}{\textbf{Subsystem}} & \multicolumn{2}{c|}{\textbf{Memory}} & \multicolumn{2}{c|}{\textbf{CPU}} & \multicolumn{2}{c|}{\textbf{Kernel}} & \multicolumn{2}{c|}{\textbf{Crash}} & \multicolumn{2}{c}{\textbf{Hang}} \\
 & Out & In & Out & In & Out & In & Out & In & Out & In \\
\hline
RAM         & 25.4 & 40.2 & 22.8 & 21.2          & 9.0           & 26.2 & 34.5          & 63.8 & 27.3          & 34.3 \\
Disk/IO     & 23.8 & 39.8             & 6.9  & 45.1 & 22.8 & 10.9             & 44.5 & 60.2             & 18.1          & 41.5 \\
Kernel      & 28.8 & 14.8    & 38.7 & 7.1  & 22.9          & 9.6              & 30.5          & 15.0             & 4.7           & 0.3 \\
CPU/Proc.   & 31.5 & 17.4             & 15.4 & 8.3           & 14.1          & 21.6             & 41.4          & 13.5             & 31.8 & 7.9 \\
Network     & 3.9  & 1.3              & 1.2  & 3.3           & 2.8           & 3.4              & 4.3           & 2.7              & 3.0           & 0.9 \\
\hline
\end{tabular}
\label{tab:granger-roles}
\end{table}

Granger causality analysis, which tests whether past values of one subsystem's telemetry improve prediction of another beyond its own history, produced mode-specific directed graphs revealing how degradation propagates across subsystems. Table \ref{tab:granger-roles} summarizes subsystem roles through aggregated out-strength (driver capacity) and in-strength (recipient exposure), exposing distinct causal signatures per failure mode.

Memory failures exhibited a classic resource exhaustion cascade with strong bidirectional interaction between RAM and Disk/IO (RAM\,$\rightarrow$\,Disk edge weight: 14.1), consistent with paging and swapping pressure. CPU failures, contrary to being self-contained, displayed strong propagation from the Kernel to the Disk/IO subsystem (weight: 18.6), indicating that high execution load disrupts kernel-level logging and I/O scheduling paths. Kernel failures showed a unique inversion where Disk/IO itself acts as a driver (out-strength: 22.8) working with the Kernel subsystem to saturate RAM (in-strength: 26.2), a propagation pattern absent from other modes. Crash failures exhibited the strongest overall causal coherence, with a tight feedback loop between Disk, RAM, and CPU producing the highest aggregated edge weights in the dataset. Hang failures were dominated by CPU/Process as the primary driver (out-strength: 31.8) with Disk/IO absorbing most of the downstream stress (in-strength: 41.5), consistent with the I/O deadlock interpretation derived from the temporal analysis.

A cross-cutting finding emerges from the recipient column of Table \ref{tab:granger-roles}, Disk/IO acts as a universal downstream sink, exhibiting disproportionately high in-strength across CPU (45.1), Crash (60.2), and Hang (41.5) modes. This suggests that the Disk/IO subsystem absorbs system stress regardless of root cause, making it a poor diagnostic discriminator despite its strong deviation signals. Conversely, the Kernel subsystem acts as an upstream initiator in CPU (out: 38.7) and Crash (out: 30.5) failures, often exerting more causal force than the CPU subsystem itself. Kernel failures (Mode 5) stand out as the only mode where this pattern inverts, with Disk/IO becoming a co-driver rather than a recipient. Edge stability analysis showed that Crash failures produced the most reproducible causal structure (14 stable edges, 100\% reproducibility on dominant loops), contradicting the expectation of chaotic collapse. Even for the smallest mode (Hang, $n=3$), 7 of 13 retained edges were significant in more than $80\%$ of runs, since $>\!80\%$ at $n=3$ requires significance in all three runs; each such edge survives leave-one-run-out removal, indicating the reported Hang structure is not driven by a single run. Median Granger lag estimates of 3-7 seconds across all modes indicate that causal propagation is gradual enough to be intercepted by predictive mechanisms.

From a practical perspective, these analyses complement the detector in two ways. ERS characterizes how early different failure modes become reliably detectable, helping calibrate expectations about attainable warning times. Granger-based subsystem roles distinguish likely upstream drivers from downstream recipients, helping practitioners interpret whether a strong signal is likely to be diagnostically informative or merely a downstream consequence of broader system stress.

\subsection{Diagnostic Explainability}

Three complementary diagnostic mechanisms were evaluated, each addressing a distinct operational question. 

\begin{table}[htbp]
\centering
\footnotesize 
\setlength{\tabcolsep}{4pt} 
\caption{Pairwise contrastive discrimination: AUC and statistical significance.}
\begin{tabular}{lcccc}
\toprule
Mode Pair & AUC & Permutation & Top & Sign \\
          &     & $p$         & Discriminator & \\
\midrule
Crash--CPU   & 0.92 & 0.012 & \texttt{apps.files.open} & + CPU \\
Crash--Hang  & 0.90 & 0.060 & \texttt{mem.kernel.Slab} & + Hang \\
CPU--Memory  & 0.97 & 0.002 & \texttt{apps.mem.logs}   & + Memory \\
CPU--Kernel  & 0.84 & 0.062 & \texttt{apps.mem.logs}   & + Kernel \\
CPU--Hang    & 1.00 & 0.007 & \texttt{mem.kernel.Slab} & + Hang \\
Memory--Hang & 1.00 & 0.005 & \texttt{mem.kernel.Slab} & + Hang \\
\bottomrule
\end{tabular}
\label{tab:contrastive_summary_up}
\end{table}

Contrastive fingerprints, which identify the features that most strongly discriminate between each pair of failure modes through signed logistic regression coefficients, achieved strong pairwise separation across most mode pairs (Table \ref{tab:contrastive_summary_up}). Comparisons involving Hang failures achieved perfect separability (AUC=1.00), driven by the distinctive kernel resource exhaustion (Slab/stack) that contrasts with user-space exhaustion in Memory or CPU modes. Crash-CPU discrimination (AUC=0.92) was driven by a contrast between memory pressure (Crash) and file descriptor/thread exhaustion (CPU). The CPU-Kernel pair exhibited moderate separation (AUC=0.84, p=0.062), reflecting partial symptom overlap. Leave-One-Run-Out (LORO) cross-validation confirmed discriminative stability across runs.

Feature-level analysis revealed diagnostically important distinctions. At the level of aggregate system metrics (e.g., total memory usage), both Hang and Memory failures present as memory exhaustion. However, the signed logistic regression coefficients from contrastive fingerprinting reveal a finer distinction, Hang failures are driven by kernel-space memory consumption (slab allocations, stack usage), while Memory failures are driven by user-space application memory (app.mem). This distinction is critical for operational response, as kernel-space and user-space exhaustion require different remediation strategies.

Rule extraction, which derives compact IF-THEN conditions from pre/post-activation telemetry changes using bagged decision trees, exhibited a strict trade-off between stability and predictive power. Memory failures were the only mode where rules triggered on the audit set (recall 0.67), though with low precision (0.26), indicating capture of generic stress markers rather than specific root causes. For CPU, Kernel, and Hang modes, rules that appeared stable during training achieved zero recall during audit, indicating that threshold-based conditions derived from training were too rigid to capture failure variability. No stable rules could be extracted for Crash failure. Although Crash failures exhibit early and consistent detection onset (ERS median 97s), their post-activation telemetry patterns are highly variable across runs, preventing the extraction of threshold-based conditions that generalize beyond the training set. 

These results demonstrate that while symbolic rules can theoretically offer deterministic human-readable explanations, the current approach provides reliable confirmation only for structurally consistent failure modes with early-onset signatures.

Counterfactual analysis, which identifies minimal feature changes that would shift a failure-indicative state toward healthy classification, varies substantially in effectiveness across failure modes. Kernel and Hang failures exhibited high controllability, probability reductions exceeding 75\% with minimal modifications (2-6 feature edits), successfully reaching the safety threshold (P$<0.3$). In contrast, Crash failures demonstrated counterfactual resistance, despite 5 targeted modifications, probability decreased by only 9.6\%. Memory and CPU failures occupied intermediate positions (29-36\% reduction) without reaching the safety threshold.

Counterfactual effectiveness exhibited an inverse relationship with causal driver strength. Modes with strong, dominant upstream drivers (Crash: high Disk/IO out-strength) proved structurally resistant to intervention, whereas modes with distributed or weaker drivers (Kernel, Hang) allowed high-leverage interventions. This \enquote{driver-resistance paradox} suggests that failures driven by resource constraints (locks, limits) offer specific keys for recovery, whereas failures driven by massive subsystem saturation are causally obvious but difficult to reverse. 

Comparative assessment confirms that no single explainability mechanism dominates across all quality dimensions. Contrastive fingerprints achieve high discriminative precision (AUC $> 0.85$) and stability ($\rho =0.81$) but provide no actionable guidance. Symbolic rules offer compactness and human readability but fail for heterogeneous modes. Counterfactual interventions deliver prescriptive guidance for resource locked modes but are ineffective against irreversible cascades. These trade-offs validate the layered diagnostic design: contrastive fingerprints for differential diagnosis, rules for confirmation where available, and counterfactuals for intervention planning.

In practice, these mechanisms serve complementary diagnostic needs: fingerprints are most useful for post-alarm mode discrimination, rules for auditable confirmation where stable patterns exist, and counterfactuals for prescriptive intervention guidance.

\subsection{Cross-Workload Validation and Generalization Boundaries}

The core approach, consensus-based detection with One-vs-Rest diagnosis, was trained exclusively on the CPU-intensive workload and then evaluated on two unseen workloads with all learned artifacts kept frozen, that is, without retraining, feature reselection, or model adaptation. Table \ref{tab:primary_validation_summary} summarizes the resulting performance across the main evaluation dimensions.

\begin{table}[htbp]
\centering
\footnotesize
\caption{Primary cross-workload validation: consensus detection and diagnosis performance.}
\begin{tabularx}{\columnwidth}{l*{4}{>{\centering\arraybackslash}X}}
\toprule
Workload & Detection Rate & Median Lead Time (sec) & False Alarm Rate & Diagnostic Accuracy (mode at alarm) \\
\midrule
NGio (I/O)     & 91.1\% (41/45) & 43 & 0.9\%  & 34.1\% (14/41) \\
NGmtr (Memory) & 94.0\% (47/50) & 83 & 0.92\% & 46.8\% (22/47) \\
\bottomrule
\end{tabularx}
\label{tab:primary_validation_summary}
\end{table}

 Cross-workload detection rates remained high (91-94\%) despite substantial telemetry distribution shifts. NGio stressed disk and I/O subsystems largely dormant during training, NGmtr activated memory bandwidth and cache hierarchy. Consensus feature representations preserved discriminative structure because degradation mechanisms remained infrastructure-invariant even as absolute metric magnitudes changed. Workload-local threshold calibration, performed using only golden-run (healthy) data from each target workload, successfully isolated distributional scale differences without requiring feature reselection or model retraining. This means that no failure or fault data from the validation workloads was used at any stage.

Lead times exhibited systematic workload dependence, NGmtr afforded nearly $2\times$ longer warning windows than NGio, reflecting intrinsic differences in failure progression dynamics under different resource regimes. This validates that early-warning potential is mechanism  dependent rather than detector dependent.

While detection generalized robustly, failure mode diagnosis proved substantially more sensitive to workload shift (34.1-46.8\% accuracy). Confusion matrix analysis revealed that misclassifications were highly structured, under NGio, Kernel and CPU failures were predominantly mapped to Memory related modes due to overlapping telemetry signatures induced by sustained I/O pressure. These patterns indicate that the classifier preserves sensitivity to failure presence but cannot fully disentangle workload-dependent manifestations. Detection decisions are triggered by consensus deviation magnitude, whereas diagnosis relies on relative feature attribution patterns more tightly coupled to workload-specific dynamics. Consensus detection consistently outperformed a subsystem aggregation baseline by $+8-12\%$ in detection rate, $+15-25s$ in lead time, and $-40\%$ in false alarm rate, validating the multi-view consensus design. 

Cross-Mode Transferability using LOMO validation established a fundamental generalization boundary. LOMO classification achieved zero accuracy across all folds (0 of 42 correct). Misclassifications were systematic: CPU$\xrightarrow{}$Crash (82\%), Memory$\xrightarrow{}$Kernel (88\%), Hang$\xrightarrow{}$Memory (100\%). Unseen modes projected onto structurally similar known modes, demonstrating that classifiers learned mode-specific discriminative combinations rather than transferable degradation abstractions.

Despite complete diagnostic failure, certain stress indicators recurred across modes (VmallocUsed, group.Iwrites, context switches), but these capture generic system stress inadequate for mode identification. Discriminative structure resides in mode-specific feature combinations and temporal orderings that cannot be reconstructed from related modes.

These results indicate a clear practical limit: the approach could not reliably diagnose failure mechanisms absent from training data. Cross-workload generalization of detection succeeds because workloads activate different resources while preserving underlying mechanisms, cross-mode generalization of diagnostic classification fails because modes correspond to qualitatively distinct degradation pathways. Effective deployment therefore requires representative training coverage of anticipated failure types plus explicit handling of novel failures as unclassified anomalies. 

To contextualize performance, three alternative strategies were compared: forecasting-based residual drift (80\% detection, 13.6\% false alarms), ERS-based instability monitoring (27-32\% detection but 170-180s lead times), and multi-detector fusion (75\% detection, 3.2\% false alarms). None outperformed the primary consensus detector, confirming that consensus aggregation provides the best sensitivity-specificity balance within a single integrated score.

These results suggest a practical division of labor among the proposed components. The consensus-based detector is suitable for cross-workload failure anticipation under modest workload recalibration using healthy data only. Diagnostic mechanisms, however, should be treated more cautiously, as their outputs are more sensitive to workload-dependent manifestations and may not transfer reliably to unseen failure modes. In deployment, this suggests using detection as the primary trigger, temporal-causal analyses to contextualize the alarm, and diagnosis as a best-effort interpretation layer rather than a universally reliable classifier.

\section{Discussion}

\subsection{Interpreting the Core Findings}

The results suggest three key insights about \gls{OFP} that are informative beyond the specific experimental setting studied here.

\paragraph{Why does consensus generalize across workloads?}
This robustness does not appear to stem from invariance in raw metric magnitudes, which change significantly across workloads, but from stability in the relational structure of degradation patterns. Features selected through consensus continued to co-deviate in characteristic ways even as their absolute scales changed. Workload-local normalization addressed nuisance variation in magnitude, while consensus preserved the ordering and co-occurrence relationships that encode failure mechanisms. Memory pressure metrics, for example, remained diagnostic regardless of whether pressure originated from computation, I/O buffering, or allocation faults, because the same kernel subsystems ultimately mediated degradation. The consensus mechanism succeeded by privileging features supported across statistical deviation, predictive utility, and structural discrimination, thereby filtering workload-specific artifacts without collapsing into overly abstract representations.

At the same time, the detection-diagnosis gap (91-94\% detection vs 34-46\% diagnostic accuracy) suggests a meaningful structural separation. Detection relies on conserved stress signatures, aggregate deviation patterns that transcend workload-specific execution dynamics. Diagnosis, by contrast, requires mode-specific feature attribution patterns that are more tightly coupled to how a particular workload exercises the OS. This separation suggests that detection and diagnosis should be treated as architecturally distinct objectives in OFP system design: detection can be workload-agnostic, but precise diagnosis may require workload-aware calibration or complementary diagnostic layers.

\paragraph{Why cross-mode transfer fails?} The complete diagnostic failure observed under LOMO validation does not appear to be simply a model-capacity limitation; rather, it suggests that the evaluated failure modes follow qualitatively distinct degradation pathways. While different modes share generic stress indicators (e.g., VmallocUsed, context switches), the discriminative structure required for diagnosis resides in mode-specific combinations, temporal orderings, and interaction patterns that cannot be reconstructed from related modes. Cross-workload generalization succeeds because workloads activate different resources while preserving underlying mechanisms. Cross-mode generalization fails because modes correspond to structurally distinct causal dynamics.

This finding supports a key design decision: explainability layers are computed independently per failure mode rather than forced into a unified embedding. This choice sacrifices zero-shot generalization but preserves diagnostic precision and interpretability. It also carries a practical deployment implication: unseen failure modes may be detectable as anomalies (via aggregate deviation) but cannot be meaningfully diagnosed without representative training data.

\paragraph{Early-warning capability is mechanism dependent} The wide variation of ERS onset times (Table \ref{tab:ers_activation}) reflects intrinsic properties of the failure mechanism rather than limitations of the detection algorithm. Failures with earlier detectable instability (e.g., CPU) afforded substantially larger warning windows, whereas modes whose signatures emerged closer to failure (e.g., Memory and Hang) offered much less time for intervention. This implies that OFP systems must adapt alerting logic to failure semantics rather than enforcing uniform operating points, a system optimized for early detection of one mode will necessarily perform differently on another. The proposed approach accommodates this by characterizing, rather than obscuring, mode-specific temporal structure.

\subsection{Comparison to Prior Work}

Many contemporary \gls{OFP} approaches emphasize predictive accuracy, often relying on models that provide limited insight into how predictions are formed. Ensemble methods combining RF or gradient-boosted trees report detection rates of 85-90\% on fault-injection datasets \cite{b1,b3}. Deep learning approaches operating on logs or multivariate telemetry report comparable or slightly higher accuracy (up to 92\%) under controlled conditions \cite{b2, b4}. However, these systems typically provide limited interpretability, relying on post-hoc feature importance without explicit temporal or causal structure.

The proposed approach achieves comparable detection performance (91-94\%) under cross-workload validation, a substantially more demanding evaluation protocol, while augmenting the predictive core with complementary analyses for interpretation and diagnosis. Rather than treating interpretability as an auxiliary concern, the approach incorporates complementary explainability-oriented analyses at multiple stages: feature selection via multi-view consensus, temporal characterization via ERS, and subsystem attribution via Granger causality. This integration delivers diagnostic insight without degrading detection performance, challenging the common assumption that interpretability necessarily trades off against accuracy. 

Temporal modeling further differentiates the framework from prior work. Several studies incorporate time-series anomaly detection \cite{b5,b6,b10}, but they typically focus on point-wise anomaly scoring rather than population-level progression analysis. ERS contributes a complementary perspective by quantifying when instability becomes reliably detectable across runs, enabling direct comparison of early-warning potential across failure modes, a perspective largely absent from prior OFP literature. 

Finally, the LOMO results provide rare empirical evidence on the limits of cross-mode generalization. Prior work often implicitly assumes transferability across failure types. Our results suggest that the discriminative structure needed for diagnosis is strongly mode-specific, refining expectations for data-driven OFP systems and underscoring the importance of representative training coverage.

\subsection{How the Components Complement Each Other in Practice}
In practice, the proposed components serve different operational roles. The consensus-based detector is suited for automated, continuous monitoring with workload-local recalibration. The remaining components function as post-alarm decision-support tools, ERS characterizes attainable warning times, Granger analysis identifies likely upstream drivers, contrastive fingerprints and rules support differential diagnosis among known modes, and counterfactuals indicate whether plausible corrective changes exist. This layered workflow automates detection while providing structured analytical evidence for interpretation and response timing.

\section{Threats to Validity}
Although this work followed a well-defined methodology, some threats to validity should be discussed. 

\subsection{Internal Validity}

The study was conducted on a single Linux kernel version, hardware configuration, and monitoring stack. This controlled the environment across workloads and failure modes, but it also limited the range of behaviors observed. Some failure modes, particularly Hang ($n=3$), are represented by few runs, which reduces statistical stability for those cases. To reduce optimistic bias, the study used LORO validation where applicable, frozen-artifact evaluation on unseen workloads, and causal streaming constraints that prevented the use of future observations at prediction time.

The dataset was generated through representative SWIFI. This provides controlled and reproducible fault activation, but the injected faults may not capture the full diversity of naturally occurring production failures. The validation workloads were selected to induce workload variation, but they remain controlled stress programs. Threshold recalibration on target workloads was performed using only golden runs rather than target fault or failure data, which reduces adaptation bias. 

\subsection{Construct Validity}

ERS was used as a measure of when instability becomes reliably detectable. It does not directly measure the underlying failure mechanism, and earlier ERS activation should not be interpreted as evidence of a more gradual failure process. Granger causality was used to characterize directed predictive relationships between subsystems. It does not establish mechanistic causation and cannot distinguish direct influence from shared latent drivers.

The diagnostic mechanisms also have construct-level limits. Contrastive fingerprints identify discriminative coefficients rather than causal explanations. Symbolic rules represent threshold-based patterns learned from the available data and may not remain valid when manifestations are heterogeneous. Counterfactuals represent model-based changes that alter the prediction, not guaranteed real-world interventions. More generally, the study uses telemetry deviation, temporal onset, subsystem interactions, and diagnostic separability as analytical proxies for failure interpretation rather than direct measurements of root cause.

\subsection{External Validity}

The study used a Linux dataset generated on a single-node environment with a specific kernel version. Production environments often include additional factors such as concurrent faults, configuration drift, multi-tenant interference, distributed dependencies, and software evolution. While this limits direct generalization to production, it does not make the findings uninformative for understanding the behavior of the approach in a controlled \gls{OS}-level setting.

To examine generalization beyond the training workload, the study used cross-workload validation with frozen artifacts, evaluating transfer to operationally distinct workloads without retraining, feature reselection, or model adaptation. This provides a stronger test of generalization than random within-workload partitioning, but it remains bounded by shared infrastructure assumptions, including the same kernel family, hardware, fault-injection process, and monitoring stack.

The evaluated failure modes do not cover all possible \gls{OS}-level failures, and the study remains bounded to Linux rather than multiple operating systems or architectural contexts. Still, the selected modes span several recurring failure categories commonly observed at the OS level, meaning they still provide evidence about the behavior of the approach across a relevant subset of failure phenomena.

\section{Conclusion}
This paper presented an explainable approach to \gls{OFP} in Linux systems that combines consensus-based detection with temporal, causal, and diagnostic analyses. Together, these analyses make it possible to examine which signals support detection, when instability becomes detectable, how degradation propagates across subsystems, and how known failure modes can be differentiated once an alarm is raised.

The study indicates that these components are complementary, but not equally effective under all conditions. Detection remained more robust to workload variation than diagnosis, and the diagnostic mechanisms differed in usefulness depending on the failure mode and the type of interpretation required.

The study also suggests an important boundary for data-driven \gls{OFP} in Linux. While predictive signals transferred across workloads with limited recalibration, diagnosis did not transfer to previously unseen failure modes. This indicates that detection and diagnosis should be treated as distinct objectives, and that representative coverage of anticipated failure types remains necessary when diagnosis is required.

Future work should extend this line of research to production environments with real failure data, develop adaptive mechanisms for handling non-stationary workloads, incorporate explicit novelty-detection pathways for unseen failure modes, and study how these analyses scale to distributed and containerized systems, as well as a systematic sensitivity analysis of the pipeline's parameters (e.g., ERS thresholds, Granger lag, persistence filtering, and correlation pruning) toward a generalizable parameterization.

\section*{Data and Code Availability}
The dataset analyzed in this work is publicly available at \url{https://github.com/jrcampos/linux-ofp-dataset}. The complete replication package pipeline implementation, analysis scripts, and the configuration required to reproduce all reported results is archived at Zenodo, DOI: \href{https://doi.org/10.5281/zenodo.21725394}{10.5281/zenodo.21725394}, and the companion input data and intermediate outputs at DOI: \href{https://doi.org/10.5281/zenodo.21630042}{10.5281/zenodo.21630042}.

\section*{Acknowledgment}

This work is funded by national funds through FCT – Foundation for Science and Technology, I.P., within the scope of the research unit UID/00326 - Centre for Informatics and Systems of the University of Coimbra, https://doi.org/10.54499/UID/00326/2025.

\vspace{12pt}

\end{document}

%% file: glossary.tex
\newacronym{ERS}{ERS}{Earliest Reliable Signal}
\newacronym{OFP}{OFP}{Online Failure Prediction}
\newacronym{OS}{OS}{Operating System}
\newacronym{XAI}{XAI}{Explainable Artificial Intelligence}

%% file: IEEE-conference-template-062824.bbl
\begin{thebibliography}{00}
\bibitem{b1} Pinciroli Vago, N. O., Forbicini, F., \& Fraternali, P. (2024). Predicting machine failures from multivariate time series: An industrial case study. Machines, 12(6), 357.
\bibitem{b2} Hadadi, F., Dawes, J. H., Shin, D., Bianculli, D., \& Briand, L. (2024). Systematic evaluation of deep learning models for log-based failure prediction. Empirical Software Engineering, 29(5), 105.
\bibitem{b3} Yadav, D. K., Kaushik, A., \& Yadav, N. (2024). Predicting machine failures using machine learning and deep learning algorithms. Sustainable manufacturing and service economics, 3, 100029.
\bibitem{b4} Nedelkoski, S., Bogatinovski, J., Acker, A., Cardoso, J., \& Kao, O. (2020, November). Self-attentive classification-based anomaly detection in unstructured logs. In 2020 IEEE international conference on data mining (ICDM) (pp. 1196-1201). IEEE.
\bibitem{b5} Blázquez-García, A., Conde, A., Mori, U., \& Lozano, J. A. (2021). A review on outlier/anomaly detection in time series data. ACM computing surveys (CSUR), 54(3), 1-33.
\bibitem{b6} Cook, A. A., Mısırlı, G., \& Fan, Z. (2019). Anomaly detection for IoT time-series data: A survey. IEEE Internet of Things Journal, 7(7), 6481-6494.
\bibitem{b7} de Campos, J. R. (2021). Advanced online failure prediction through machine learning (Doctoral dissertation, Universidade de Coimbra (Portugal)).
\bibitem{b8} Campos, J. R., Costa, E., \& Vieira, M. (2023, October). Online failure prediction through fault injection and machine learning: Methodology and case study. In 2023 IEEE 34th International Symposium on Software Reliability Engineering (ISSRE) (pp. 451-461). IEEE.
\bibitem{b9} Campos, J. R., \& Nogueira, R. P. (2024, May). Statistical Process Control for Supporting OS-level Failure Prediction. In Workshop de Testes e Tolerância a Falhas (WTF) (pp. 99-103). SBC.
\bibitem{b10} Ren, H., Xu, B., Wang, Y., Yi, C., Huang, C., Kou, X., ... \& Zhang, Q. (2019, July). Time-series anomaly detection service at microsoft. In Proceedings of the 25th ACM SIGKDD international conference on knowledge discovery \& data mining (pp. 3009-3017).
\bibitem{b16} Ali, S., Abuhmed, T., El-Sappagh, S., Muhammad, K., Alonso-Moral, J. M., Confalonieri, R., ... \& Herrera, F. (2023). Explainable Artificial Intelligence (XAI): What we know and what is left to attain Trustworthy Artificial Intelligence. Information fusion, 99, 101805.
\bibitem{b17} Saranya, A., \& Subhashini, R. (2023). A systematic review of Explainable Artificial Intelligence models and applications: Recent developments and future trends. Decision analytics journal, 7, 100230.
\bibitem{b18} Liu, Y., \& Jafarpour, B. (2024). Graph attention network with Granger causality map for fault detection and root cause diagnosis. Computers \& Chemical Engineering, 180, 108453.
\bibitem{b19} Wang, J. G., Chen, R., Ye, X. Y., Xie, Z. T., Yao, Y., \& Liu, L. L. (2024). A hierarchical granger causality analysis framework based on information of redundancy for root cause diagnosis of process disturbances. Computers \& Chemical Engineering, 182, 108589.
\bibitem{b20} Wachter, S., Mittelstadt, B., \& Russell, C. (2017). Counterfactual explanations without opening the black box: Automated decisions and the GDPR. Harv. JL \& Tech., 31, 841.

\bibitem{zhang2020minority} Zhang, J., Zhou, K., Huang, P., He, X., Xie, M., Cheng, B., Ji, Y., \& Wang, Y. (2020). Minority disk failure prediction based on transfer learning in large data centers of heterogeneous disk systems. IEEE Transactions on Parallel and Distributed Systems.

\bibitem{wang2017failure} Wang, Z., Zhang, M., Wang, D., Song, C., Liu, M., Li, J., Lou, L., \& Liu, Z. (2017). Failure prediction using machine learning and time series in optical network. Optics Express, 25(16), 18553--18565.

\bibitem{chen2014failure} Chen, X., Lu, C.-D., \& Pattabiraman, K. (2014). Failure prediction of jobs in compute clouds: A Google cluster case study. In 2014 IEEE International Symposium on Software Reliability Engineering Workshops (pp. 341--346). IEEE.

\bibitem{campos2020} Campos, J. R., Vieira, M., \& Costa, E. (2020). Fault injection to generate failure data for failure prediction: A case study. In 2020 IEEE 31st International Symposium on Software Reliability Engineering (ISSRE). IEEE.

\bibitem{salfner2010} Salfner, F., Lenk, M., \& Malek, M. (2010). A survey of online failure prediction methods. ACM Computing Surveys (CSUR), 42(3), 1-42.

\bibitem{jassas2018} Jassas, M., \& Mahmoud, Q. H. (2018, October). Failure analysis and characterization of scheduling jobs in google cluster trace. In IECON 2018-44th Annual Conference of the IEEE Industrial Electronics Society (pp. 3102-3107). IEEE.

\bibitem{granger} Granger, C. W. (1969). Investigating causal relations by econometric models and cross-spectral methods. Econometrica: journal of the Econometric Society, 424-438.

\bibitem{benjamini1995} Benjamini, Y., \& Hochberg, Y. (1995). Controlling the false discovery rate: a practical and powerful approach to multiple testing. Journal of the Royal statistical society: series B (Methodological), 57(1), 289-300.

\end{thebibliography}
